\documentclass[conference, 10pt]{IEEEtran}

\usepackage{cite}
\usepackage{url}
\usepackage{bm}
\usepackage{graphicx}
\usepackage{color}
\usepackage{placeins}
\usepackage{float}
\usepackage{tabularx,colortbl}
\usepackage{ifthen}
\usepackage{amsmath,amsfonts}
\usepackage[caption=false]{subfig}

\makeatletter

\newcounter{author}
\renewcommand{\author}[2][]{
   \stepcounter{author}
   \@namedef{author@\theauthor}{#2}
   \@namedef{authorlabel@\theauthor}{#1}
}

\newcounter{address}
\newcommand{\address}[2][]{
   \stepcounter{address}
   \@namedef{address@\theaddress}{#2}
   \@namedef{addresslabel@\theaddress}{#1}
}

\newcommand{\alsep}{and}

\def\newmaketitle{\par%
  \begingroup%
  \normalfont%
  \def\thefootnote{}
  \def\footnotemark{}
  \let\@makefnmark\relax
  \footnotesize
  \footnotesep 0.7\baselineskip
  \normalsize%
  \twocolumn[\thenewmaketitle\@IEEEaftertitletext]%
  \if@IEEEusingpubid
     \enlargethispage{-\@IEEEpubidpullup}%
  \fi
  \endgroup
  \setcounter{footnote}{0}\let\maketitle\relax\let\@maketitle\relax
  \gdef\@thanks{}%
  \let\thanks\relax}

\def\thenewmaketitle{
  \newpage
  \begin{center}%
    \vskip0.2em{\Huge\@IEEEcompsoconly{\sffamily}\@IEEEcompsocconfonly{\normalfont\normalsize\vskip 2\@IEEEnormalsizeunitybaselineskip
   \bfseries\large}\@title\par}\vskip1.0em\par%
    \vspace{1ex}
    \newcounter{c@author}
    \newcounter{c@tmp}
    \ifthenelse{\value{author}=2}{%
      \newcommand{\liand}{ and }}{%
      \newcommand{\liand}{, and }}
    \ifthenelse{\value{address}<2}{%
      \@nameuse{author@1}%
      \stepcounter{c@author}%
      \whiledo{\value{c@author}<\value{author}}{%
        \setcounter{c@tmp}{\value{author}}%
        \addtocounter{c@tmp}{-\value{c@author}}%
        \ifthenelse{\value{c@tmp}=1}{%
          \renewcommand{\alsep}{\liand}}{\renewcommand{\alsep}{, }}%
        \stepcounter{c@author}\alsep \@nameuse{author@\thec@author}}\\%
    }
    {
      \@nameuse{author@1}${}^{(\ref{\@nameuse{authorlabel@1}})}$%
      \stepcounter{c@author}%
      \whiledo{\value{c@author}<\value{author}}{%
      \setcounter{c@tmp}{\value{author}}%
      \addtocounter{c@tmp}{-\value{c@author}}%
      \ifthenelse{\value{c@tmp}=1}{%
        \renewcommand{\alsep}{\liand}}{\renewcommand{\alsep}{, }}%
      \stepcounter{c@author}\alsep \@nameuse{author@\thec@author}%
        ${}^{(\ref{\@nameuse{authorlabel@\thec@author}})}$%
      }
    }
    \vspace{0.2ex}

    \ifthenelse{\value{address}>0}{%
      \ifthenelse{\value{address}=1}{
        {\@nameuse{address@1}}
      }
      {
        \newcounter{c@address}

        \begin{center}
        \whiledo{\value{c@address}<\value{address}}
        {
          \refstepcounter{c@address}
            ${}^{(\thec@address)}$\,%
              \label{\@nameuse{addresslabel@\thec@address}}%
              \@nameuse{address@\thec@address}\\ %
        }
        \end{center}
      } 
    }
    {
      \relax
    }
  \end{center}
}

\makeatother

\title{Inverse Design of Frequency Selective Surface Using Physics-Informed Neural Networks}

\author{Yu-Hang Liu}
\author{Bing-Zhong Wang}
\author{Ren Wang}

\address{Institute of Applied Physics, University of Electronic Science and Technology of China
,Chengdu, 611731, China (bzwang@uestc.edu.cn)}

\begin{document}

\newmaketitle

\begin{abstract}
This paper uses Physics-Informed Neural Network (PINN) to design Frequency Selective Surface (FSS). PINN integrates physical information into the loss function, so training PINN does not require a dataset, which will be faster than traditional neural networks for inverse design. The specific implementation process of this paper is to construct a PINN using field solutions of mode matching method, and given the design goal, the PINN can train the shape of the diaphragm. The single frequency FSS that meets the design goal was designed using the inverse design method proposed in this paper without a dataset, verifying the rationality of using PINN to design metasurfaces. Using PINN for inverse design is not limited to single frequency FSS, but can also be used for more complex metasurface.
\end{abstract}

\section{Introduction}
In 2019, Raissi et al.{\cite{pinn1}} first proposed PINN, which has since attracted widespread attention. PINN achieves the goal of training without a dataset by incorporating physical information into the loss function. PINN was first developed in fluid mechanics, for example, Mehta et al.{\cite{pinn2}} obtained a variable order fractional model using PINN to predict the mean velocity distribution and Reynolds stress at any Reynolds number. Gradually, PINN also began to be applied in the electromagnetics{\cite{pinn3,pinn4}}. Chen and Lu et al. used PINN for metamaterial design to solve the problem of inverse scattering of dielectric{\cite{pinn5}}. In addition, Chen et al. also used PINN to solve the inverse design problem of the shape of scatterers in electromagnetic devices{\cite{pinn6}}. These studies indicate that PINN can be used for inverse design of electromagnetic structures, but current inverse designs are all dielectric-loaded and there have been few issues with metal-loaded.

Based on the advantages of PINN, it can be applied in inverse design. In this paper, PINN is used for inverse design of single frequency FSS. In Section II, a method for inverse designing FSS using PINN is proposed. In Section III, the training results are presented and simulated. In Section IV, analysis and outlook are conducted on this method.

\section{Method}

The designed FSS consists of FSS unit cells arranged tightly to form an infinite rectangular array, and the unit cell is shown in Figure 1. The structure of the unit cell consists of one layer of dielectric and two layers of diaphragm. The diaphragms are Perfect Electric Conductor (PEC) with negligible thickness, attached to both sides of the dielectric. The thickness of the medium in the unit is $d$, and the relative dielectric constant is $\varepsilon_r$, the cross-section is a square with a side length of $a$.

\begin{figure}[ht]
\begin{center}
\noindent
  \includegraphics[width=2.8in]{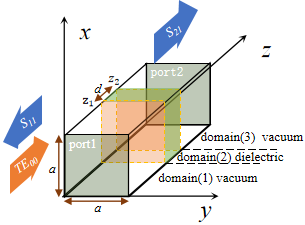}
  \caption{FSS unit cell in infinite arrays.}\label{Fig1Label}
  \vspace{-1.5em}
\end{center}
\end{figure}

The four sides of the unit cell satisfy periodic boundary conditions, and the form of its electromagnetic field solution can be described by the Floquet mode.

On the domain $x\in\left(0,a\right), y\in(0,a)$, the shape function of the two-layer diaphragms is:
\begin{equation}
    g_i(x,y) =\begin{cases}1,&\text{Vacuum,$z=z_i$}\\0,&\text{PEC,$z=z_i$}\end{cases},\\i=1,2
\end{equation}

The incident wave is $\rm{TE_{00}}$. The mode matching method is used on both sides of the metal diaphragms to establish the interaction relationship between the metal diaphragm and the electromagnetic fields. Equation (2) shows the principle of the field solution of the model.
\begin{equation}
    \sum C_{i}(\boldsymbol{E_{t(i)}},\boldsymbol{H_{t(i)}})=g\bigodot\sum C_{i+1}(\boldsymbol{E_{t(i+1)}},\boldsymbol{H_{t(i+1)}})
\end{equation}
where, $\sum$ applies to modes. $C_{i}$ represent the mode coefficients of the ($i$-th) region, $\boldsymbol{{E}_{t}}$ and $\boldsymbol{{H}_{t}}$ represent the tangential electromagnetic fields in Floquet mode. $\bigodot$ represents the interaction relationship between the diaphragms and the electromagnetic fields: when $g=0$, Equation (2) satisfies that the tangential electric field is $0$; when $g=1$, Equation (2) satisfies the boundary condition that the tangential electromagnetic fields are continuous at the interface of the dielectric and vacuum.

Usually, the inverse design goal of FSS can be represented by S parameters, which can be represented by certain mode coefficients. The control equation, boundary conditions, and design goal of the inverse design problem are already included in Equation (2), and the loss function $L$ can be directly constructed using the 2-norm of the residual in (2).
\begin{equation}
    L=\vert\vert res(Eq.(2))\vert\vert _{2}^{2}
\end{equation}

Only $g_i\left(x,y\right)$ in $L$ has not been solved. Using a fully connected neural network (FCNN) to generate the structure of the metal diaphragms $g_i\left(x,y\right)$, obtaining $res(Eq.(2))$ by interacting $g_i\left(x,y\right)$ with the electromagnetic field through operator $\bigodot$, the final PINN model is shown in Figure 2.
\vspace{-1em}
\begin{figure}[ht]
\begin{center}
\noindent
  \includegraphics[width=2.8in]{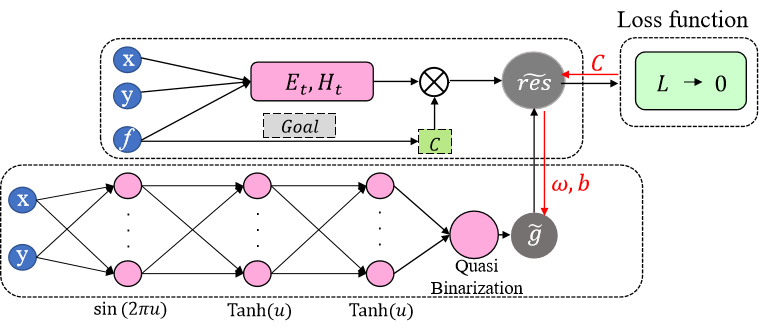}
  \caption{PINN model.}\label{Fig1Label}
  \vspace{-1.5em}
\end{center}
\end{figure}

After setting the design goal, the backpropagation algorithm is used to correct the mode coefficients and parameters in the FCNN until the loss function approaches 0, and then the $g_i\left(x,y\right)$ obtained is the final shape of the diaphragms.
\section{Example}
Using the PINN model in Section II, an FSS is trained to meet the design goal.
\subsection{Train}
The design goal is to not transmit electromagnetic waves at a frequency of 15GHz, set $\left|S_{21}\left(12GHz\right)\right|=0.99$, $\left|S_{21}\left(15GHz\right)\right|=0.1$, $\left|S_{21}\left(18GHz\right)\right|=0.99$ for $\rm{{TE}_{00}}$ mode at Port 2. In the Equation (2), 121 TE modes and 121 TM modes are used. For dielectric, $d=2mm$, $\varepsilon_{r}=3.2$, $a=10mm$. In the FCNN, there are 3 hidden layers, and each hidden layer has 32 neurons. The optimizer is Adam.

Set the training steps to $10k$, accelerate by GPU (A800), and complete the training in one hour, four minutes, and twelve seconds. Finally, the shape of the FSS diaphragms is shown in Figure 3.
\begin{figure}[htb]
    \centering
    \setlength{\fboxsep}{0pt} 
    \begin{minipage}{0.5\textwidth}
    \centering
    \subfloat[]    {\fbox{\includegraphics[width=0.3\columnwidth]{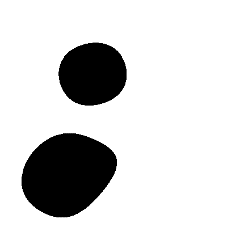}}}{\quad}
    \subfloat[]    {\fbox{\includegraphics[width=0.3\columnwidth]{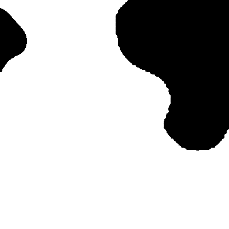}}}{\quad}
    \end{minipage}\begin{minipage}{0.5\textwidth}
    \hspace{-3.5em}
    \includegraphics[width=0.1\columnwidth]{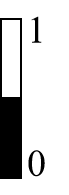}
    \end{minipage}
    \caption{The shape of FSS diaphragms. 
        (a) $g_1\left(x,y\right)$;
        (b) $g_2\left(x,y\right)$.}
    \label{fig_dipole}
    \vspace{-1.5em}
\end{figure}
\subsection{Simulation}
Simulate the structure using the frequency domain solver of full wave simulation software, and the $S_{21}$ at Port 2 was obtained as shown in Figure 4. 

Define the error as Equation (4):
\begin{equation}
    \epsilon=\frac{f_{train}-f_{goal}}{f_{goal}}\times \% 
\end{equation}

From Figure 4, it can be seen that the frequency corresponding to the minimum value of $S_{21}$ for the training model is 15.288GHz, indicating that $\epsilon=1.92\%$. The training model can meet the requirements of the design goal very well, and thus we have successfully used the PINN method to inverse design the FSS that matches the design goal.
\vspace{-1em}
\begin{figure}[ht]
\begin{center}
\noindent
  \includegraphics[width=2.8in]{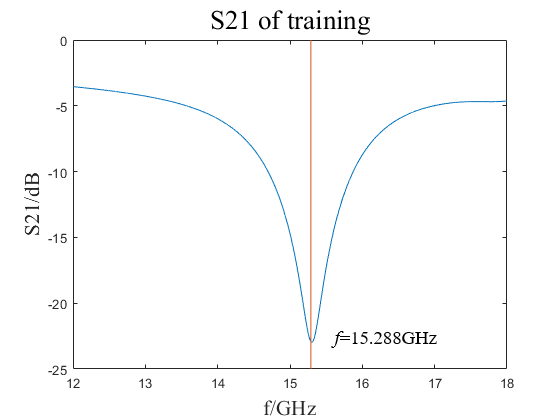}
  \caption{$S_{21}$ of training FSS.}\label{Fig1Label}
\vspace{-1.5em}
\end{center}
\end{figure}
\section{Conclusion}
Using PINN, this paper successfully designed the FSS that meets the set inverse design goal. Due to the fact that PINN does not require a dataset, it is more efficient than using traditional neural networks for inverse design. This paper introduces PINN into the design of FSS for the first time, and numerical experiment have proven the correctness of this method. Although this paper takes FSS as an example, this method is not limited to FSS. In fact, many electromagnetic devices can be designed using this method, which will promote the development of inverse design of electromagnetic devices.

\section*{ACKNOWLEDGEMENT}
This work was supported in part by the National Natural Science Foundation of China under Grants 62171081 and U2341207, the Natural Science Foundation of Sichuan Province under Grant 2022NSFSC0039, and the Aeronautical Science Foundation of China under Grant 2023Z06208002.



%
\bibliographystyle{IEEEtran}
\bibliography{refs}

\end{document}